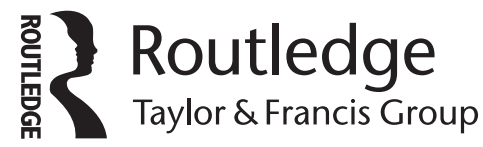

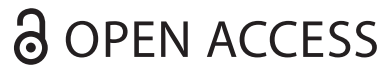

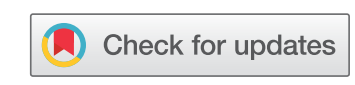

# The young and the old, the fast and the slow: a large-scale study of productivity classes and rank advancement

Marek Kwiek [a,b] and Wojciech Roszka [a,c]

[a]Center for Public Policy Studies, AMU University of Poznan, Poznan, Poland; [b]German Center for Higher Education Research and Science Studies (DZHW), Berlin, Germany; [c]Poznan University of Economics and Business, Poznan, Poland

**ABSTRACT**
We examined a large sample of Polish science, technology, engineering, mathematics, and medicine (STEMM) scientists ($N = 16{,}083$) to study rank advancement and productivity in the past 40 years. We used two previously neglected time dimensions – promotion age and promotion speed – to construct individual lifetime biographical and publication profiles. We followed a classificatory approach and the new methodological approach of journal prestige–normalized productivity. All scientists were allocated to different productivity, promotion age, and promotion speed classes (top 20%, middle 60%, and bottom 20%). The patterns found were consistent across all disciplines: scientists in young promotion age classes (and fast promotion speed classes) in the past were currently the most productive. In contrast, scientists in old promotion age classes (and slow promotion speed classes) in the past were currently the least productive. In the three largest disciplines, the young-old promotion age productivity differential for associate professors was 100–200% (150–200% for full professors), and the fast-slow promotion speed productivity differential for associate professors was 80–150% (100–170% for full professors). Our results were partly supported by a regression analysis in which we examined odds ratio estimates of belonging to top productivity classes. To examine the sample, we combined biographical and demographic data collected from the national register of all Polish scientists and publication metadata on all Polish articles indexed in Scopus ($N = 935{,}167$).

## Introduction

The relationship between promotions in academic careers and research productivity has long been examined in the literature (Bayer and Dutton 1977; Cole 1979; Long, Allison, and McGinnis 1993; Tien and Blackburn 1996). Promotions to each successive stage in an academic career occur at a certain age and after a certain period has elapsed following an earlier promotion. Thus, two underexplored independent time dimensions may affect successful (and unsuccessful) careers: promotion age and promotion speed. The impact of these two variables may vary from country to country and over time,

**CONTACT** Marek Kwiek kwiekm@amu.edu.pl Center for Public Policy Studies, Adam Mickiewicz University, ul. Szamarzewskiego 89c, 60-568 Poznan, Poland and this research is empirically driven by longitudinal data from the Polish higher education system over the past 40 years.

Sociologists of science have long argued that 'scientists are not only concerned with achieving high rank, but in doing so as quickly as possible. Some measure of recognition is garnered by the distinction of being a 'young' associate or a 'young' full professor' (Cole and Cole 1973, 130). The timing of promotions can be considered indicative of success, with variations in timing being a critical component of research on rank advancement and productivity (Long, Allison, and McGinnis 1993, 704). The faculty rank system is intended to have a motivating effect on productivity; in very general terms, publishing is reinforced by promotions, and scientists are often reported to write for promotions (Tien and Blackburn 1996, 2). Moving up the academic career ladder is a common aspiration in the Polish academe; scientists' research opportunities, academic work portfolio, working time distribution, and participation in university governance are hugely determined by their current location on the academic ladder.

Advancement in rank is a form of peer recognition. The success and failure of rank advancement in the Polish case is a proxy for success and failure in academic science. Rank advancement can be examined with the data available using biological age at promotion times and the time spent between subsequent promotions (or time in rank, Bayer and Dutton 1977, 263–4; Long, Allison, and McGinnis 1993, 705–7; Tien and Blackburn 1996, 6–8). Some Polish scientists receive successive promotions at a young age and others at an older age. Additionally, some are promoted quickly, while others are promoted slowly (time between promotions). In Poland, promotions are determined exclusively by research output, and at each career stage, publications need to be rigorously assessed by committees formed by academic peers at the national (rather than institutional) level. The time factor is not important – (internationally indexed) research achievements are, especially in STEMM disciplines on which this study is focused.

We posed the following three research questions:

(1) What is the relationship between current individual productivity and past promotion age?
(2) What is the relationship between current individual productivity and past promotion speed?
(3) What is the relationship between past promotion age (and past promotion speed) and current membership in the top productivity classes, based on the combined effects of other variables (and using regression analysis)?

We explored the career histories of 16,083 scientists with at least doctoral degrees to inquire about the patterns that have not been explored before on a similar national scale, using a combination of structured Big Data (of the Scopus type: $N = 935{,}167$ articles, 1973–2021) and biographical and demographic data (of the national registry of scientists type). To the best of our knowledge, our research is the first large-scale country-level examination of the relationships between rank advancement and productivity in the entire internationally visible STEMM national academic community over a period of four decades.

## Theoretical framework

As classical students of social stratification in science have described, 'the critical part of recognition associated with rank is the achievement of high rank in a high-prestige department at a relatively early age' (Cole and Cole 1973, 131). Abramo, D'Angelo, and Murgia (2016, 15) have recently shown that 'an individual who gains promotion to full professor at young age then maintains and increases his/her productivity more than colleagues who are promoted at later age.' Furthermore, 'the post-promotion productivity of full professors who are nominated at a young age is higher than that of their colleagues who are promoted at a later age' (Abramo, D'Angelo, and Murgia 2016, 15). The association between productivity and rank advancement has long been a scholarly theme in the sociology of science and the economics of science (e.g., David 1994; Long, Allison, and McGinnis 1993; Stephan 2012; Stephan and Levin 1992). Ranks stratify the academic profession (Zuckerman 1988), and each career step represented in subsequent promotions results in upgrades in prestige and salaries (Tien and Blackburn 1996). In the present study, scientists from the young and old promotion age classes in the past and scientists from the fast and slow promotion speed classes in the past are compared in terms of their current four-year (2014–17) productivity.

Recent changes to the academic profession have been widely documented. A particularly large strand of research is based on cross-national comparative survey designs (see Arimoto et al. 2015; Cummings and Finkelstein 2012; Fumasoli, Goastellec, and Kehm 2015; Kwiek 2019; Postiglione and Jung 2017; Teichler, Arimoto, and Cummings 2013; Teichler and Höhle 2013). These large-scale comparative empirical studies (and many others) show that ever-expanding academic science, combined with an abundance of opportunities for newcomers to the higher education sector, known from the 1960s, changed into the science of resource constraints and permanent austerity, with huge numbers of postdocs and shrinking opportunities for their first full-time employment (Finkelstein, Conley, and Schuster 2016, 99–102; Wang and Barabási 2021, 169). However, as strongly as ever before, 'recognition is key in science' (Stephan 2012, 19). This research is built on the assumption held in the Polish STEMM science system that recognition comes from publication-driven promotions to higher ranks, opening new career opportunities, including access to infrastructure and research funding.

Both biological age at being promoted (promotion age) and time spent between ranks (promotion speed) were factors in perceptions of success and failure in an academic career (Cole and Cole 1973, 130; Long, Allison, and McGinnis 1993, 704), as shown in the traditional sociology of science. Full professorship was the ultimate career goal, to which many aspired but few achieved (Hermanowicz 2012). Among the chosen few, the amount of time needed for full professorship differs by gender and discipline (Teelken, Taminiau, and Rosenmöller 2021; Mantai and Marrone 2023).

The questions of how academic promotions are associated with productivity and whether promotion age and promotion speed are differentiating factors regarding productivity are linked to wider issues of inequality in science: the distribution of productivity among scientists – in departments, institutions, and countries – is highly unequal, with a few publishing a lot and many publishing little or nothing. A defining characteristic of science is 'extreme inequality in the allocation of rewards' (Stephan 2012, 31). Inequality has been explained according to the cumulative advantage theory and the 'sacred spark' hypothesis (David 1994; Fox 1983), among others.

First, the implications for productivity in cumulative advantage theory and the accompanying reinforcement theory are clear (Stephan and Levin 1992, 29): 'Scientists productive in an early period are productive in later periods; those not productive at an early date are less likely to be productive at a later date.' In another formulation, 'productive scientists are likely to be even more productive in the future, while scientists who produce little original work are likely to decline further in their productivity' (Allison and Stewart 1974, 596).

An 'initial success' in publishing entails increasing productivity; a 'bad start' may subsequently lead to quitting research altogether (Turner and Mairesse 2005, 3). Full professors nominated at a young age show higher post-promotion productivity than peers promoted at an older age (Turner and Mairesse 2005, 17). Each step rewards subsequent research successes with better access to the means for future research successes (David 1994, 12). However, in the specific Polish case, recognition is not publication; recognition is publication-driven promotions. Early recognition makes resources available, while late recognition inhibits access to resources and reduces the chances for future productivity.

Second, according to the 'sacred spark' hypothesis, 'there are substantial, predetermined differences among scientists in their ability and motivation to do creative scientific research' (Allison and Stewart 1974, 596). Scientists comprise a heterogeneous population that contains a separate class of 'rare individuals of great talent' (David 1994, 12). Indeed, 'those with the spark always are productive. Those without it, however, never see their careers take off and flourish' (Stephan and Levin 1992, 30). The hypothesis emphasizes that a differential distribution of talent among scientists affects inequality in scientific production more than the way in which recognition is awarded in science.

## Recognition, publications, and academic ranks: the Polish case

Recognition in Poland – in STEMM disciplines and in the decades studied – was achieved through higher ranks only, and higher ranks were achieved through high-quality publications only, hence the pivotal role of productivity or high-quality publications within a unit of time. Therefore, publication-driven ranks determined who was successful in science and who was not. These unique characteristics form a specific Polish 'university configuration,' as Musselin (2010, 207) would call it.

As publication-driven rank advancement is key to academic success, early success tends to be understood as promotion at a young age, which accelerates a career, and late success tends to be understood as promotion at an old age, which hinders a career. Research in Poland tends to be conducted for recognition stemming from publication-driven rank promotions. All scientists in our sample are already assistant professors so rank advancement in their case means becoming (at some point, if applicable) associate professors and full professors.

In Poland in 2021/2022, there were 130 public universities (and 217 private universities), with 1,218,200 students, 27,661 doctoral students, and 88,416 full-time employed academics (GUS 2022). The Polish academic career structure is built around three major scientific degrees: doctorate, habilitation (or the postdoctoral degree), and full professorship. The system is highly stratified, with about 10% of full professors at the top and about 20% of associate professors below (10.17% full professors, 19.57% associate professors, 44.09% assistant professors, 14.56% lecturers and 11.61% other staff). Promotions to associate and full professorships are governed nationally (rather than institutionally) and there are no limits in the numbers of new associate and full professors at either institutional or national levels. Full professorships are awarded on the basis of successfully passing rigorous, research-based national-level promotion procedures. From a historical perspective of the past 30 years, there has been no major changes in requirements for promotions: they have always been strictly publication-related. Academics in STEMM with national publication profiles have had limited chances for promotions. Both private and public universities follow the same regulations regarding rank advancement. This research follows the terminology used by the national statistics in reporting data on academic personnel. As an analytical approach, we chose a three-degree system rather than a multiple-rank system (with 'university professors', 'ordinary professors', and 'professors without habilitation') because the latter system is not consistently applied across all institutional types.

Institutional-level 'micro-political practices' related to promotion (Teelken, Taminiau, and Rosenmöller 2021) tend to play a small role compared with systems with institutional limits of promotions. The association between academic promotion and research productivity is strong, and the research output presented for rigorous peer assessment is the single most important component of applications for promotion. Although direct promotion rewards (e.g., prestige and pay) are the same across the Polish system, including discipline and gender, indirect promotion rewards vary between scientists belonging to the young and the old promotion age classes, such as in grant application success and failure. Early promotions in academic careers, as shown in the academic vitae, generally strengthen grant applications, while late promotions generally weaken them.

Thus, in Poland, 'achieving early distinction' matters (David 1994, 19), especially under conditions of positive and negative feedback mechanisms. There is positive feedback between achievements (here, academic promotions and publications) and access to research funding. Promotions function as constructed scientific reputations: 'the fundamental currency of the reward system' (David 1994, 19). In the Polish case (different to, for example, the United States), the individual timing of promotions acts as a major differentiating factor – a major marker of success in academic careers. In STEMM, early promotions tend to suggest successful, high-quality researchers.

## Data and methods

In this research, we used a classificatory approach to academic careers. We used three parallel classification systems: (1) research productivity, (2) promotion age, and (3) promotion speed. Classifications were tripartite, with scientists ranked according to the 20/60/20 (top/middle/bottom) formula separately within each discipline. Each scientist in our sample was classified as belonging to the top, middle, or bottom class in terms of productivity; young, middle, or old class in terms of promotion age; and fast, typical, or slow class in terms of promotion speed.

First, using publication metadata from Scopus, we allocated all scientists in our sample to the top, middle, or bottom productivity classes for the 2014–17 period. Then, using the data on biological age at promotions (doctorate, habilitation, and full professorship), we allocated all scientists to different classes of promotion age. According to the tripartite formula, the class of young scientists included the upper 20% of scientists in terms of biological age at the three promotion times, and the class of old scientists included the bottom 20% of scientists in terms of biological age at these three promotion times.

Finally, again using the data on biological age and age at promotion, we allocated all scientists to different classes of promotion speeds. The class of fast scientists included the top 20% of scientists in terms of time spent between subsequent promotions, and the class of slow scientists included the bottom 20% of scientists in terms of time spent between subsequent promotions. The three classification systems were applied to each of the three career stages; scientists were classified retrospectively (i.e. when they were assistant and associate professors, if applicable). Half a century of Scopus metadata on Polish publications combined with demographic data on all Polish scientists allowed us to construct the major classifications retrospectively, comparing scientists at different career stages in specific STEMM disciplines with their exact peers.

## Dataset, sample, and variables

Data were collected from two sources: (1) The Polish Science Observatory, which is a database created and maintained by the authors, and (2) the Scopus raw data provided by the International Center for the Studies of Research Lab (ICSR Lab), including metadata on all articles published from 1973 to 2021 by authors with Polish affiliations ($N = 935{,}167$ articles). We selected only scientists in 12 STEMM disciplines ($N = 16{,}083$). All scientists in our sample had at least a doctorate, were employed full-time in higher education, and had published at least one Scopus-indexed article (the list of disciplines is provided under Table 1). The Observatory maintained by the authors covers all Polish universities with internationally visible publications within the decade of 2007–17. Its initial construction, with main steps in merging the biographical and administrative datasets using both deterministic and probabilistic approaches, was described in detail in (Kwiek and Roszka 2021a, 1350–1 and Kwiek and Roszka 2021b, 4–7). Figure 1 shows the distribution of the sample by biological age. The academic hierarchy was reflected in the age distribution. The mean ages were as follows: assistant professors, 41.0 years; associate professors, 50.8 years; and full professors, 61.7 years.

Our dataset included gender in binary form (male or female) and year of birth. We obtained the year of the first publication, which allowed us to calculate academic age or the number of years since the first publication indexed in Scopus using the application programming interface protocol. We collected all Scopus-indexed publications (type: article) in individual lifetime publication profiles for every scientist and determined the predominant discipline (the modal value) for each scientist as the most often occurring value. The description of variables is provided in Table 1 in Supplementary Material.

Three Polish academic degrees were used as proxies of internationally comparable academic positions: working with a doctoral degree only, which was regarded as a proxy for working at the rank of assistant professor ($N = 9084$); working with a habilitation or postdoctoral degree, which was

**Table 1.** Structure of the sample: all Polish university professors in STEMM with at least a single Scopus-indexed article and with at least a doctoral degree by gender, age group, academic position, and discipline.

| | Female | | | Male | | | Total | | |
|---|---|---|---|---|---|---|---|---|---|
| | *N* | % col | % row | *N* | % col | % row | *N* | % col | % row |
| Age group | | | | | | | | | |
| 39 and less | 2180 | 34.0 | 46.4 | 2516 | 26.0 | 53.6 | 4696 | 29.2 | 100.0 |
| 40–54 | 3094 | 48.2 | 42.0 | 4277 | 44.2 | 58.0 | 7371 | 45.8 | 100.0 |
| 55 and older | 1139 | 17.8 | 28.4 | 2877 | 29.8 | 71.6 | 4016 | 25.0 | 100.0 |
| Academic position | | | | | | | | | |
| Assistant Prof. | 4148 | 64.7 | 45.7 | 4936 | 51.0 | 54.3 | 9084 | 56.5 | 100.0 |
| Assoc. Prof. | 1725 | 26.9 | 36.6 | 2990 | 30.9 | 63.4 | 4715 | 29.3 | 100.0 |
| Full Professor | 540 | 8.4 | 23.6 | 1744 | 18.0 | 76.4 | 2284 | 14.2 | 100.0 |
| Discipline | | | | | | | | | |
| AGRI | 1130 | 17.6 | 53.7 | 976 | 10.1 | 46.3 | 2106 | 13.1 | 100.0 |
| BIO | 865 | 13.5 | 61.8 | 535 | 5.5 | 38.2 | 1400 | 8.7 | 100.0 |
| CHEM | 595 | 9.3 | 50.6 | 581 | 6.0 | 49.4 | 1176 | 7.3 | 100.0 |
| CHEMENG | 135. | 2.1 | 39.0 | 211 | 2.2 | 61.0 | 346 | 2.2 | 100.0 |
| COMP | 126 | 2.0 | 16.3 | 645 | 6.7 | 83.7 | 771 | 4.8 | 100.0 |
| EARTH | 284 | 4.4 | 33.7 | 559 | 5.8 | 66.3 | 843 | 5.2 | 100.0 |
| ENG | 380 | 5.9 | 14.8 | 2181 | 22.6 | 85.2 | 2561 | 15.9 | 100.0 |
| ENVIR | 685 | 10.7 | 51.9 | 635 | 6.6 | 48.1 | 1320 | 8.2 | 100.0 |
| MATER | 397 | 6.2 | 32.9 | 808 | 8.4 | 67.1 | 1205 | 7.5 | 100.0 |
| MATH | 195 | 3.0 | 25.2 | 580 | 6.0 | 74.8 | 775 | 4.8 | 100.0 |
| MED | 1478 | 23.0 | 54.5 | 1233 | 12.8 | 45.5 | 2711 | 16.9 | 100.0 |
| PHYS | 143 | 2.2 | 16.5 | 726 | 7.5 | 83.5 | 869 | 5.4 | 100.0 |
| **Total** | 6413 | 100.0 | 39.9 | 9670 | 100.0 | 60.1 | 16,083 | 100.0 | 100.0 |

Note: Twelve STEMM disciplines examined: AGRI, agricultural and biological sciences; BIO, biochemistry, genetics, and molecular biology; CHEMENG, chemical engineering; CHEM, chemistry; COMP, computer science; EARTH, earth and planetary sciences; ENG, engineering; ENVIR, environmental science; MATER, materials science; MATH, mathematics; MED medicine, and PHYS, physics and astronomy.

regarded as a proxy for working at the rank of associate professor ($N = 4715$); and working with a professorship title, which was regarded as a proxy for working at the rank of full professor ($N = 2284$). The three dates in our dataset, that is, the years in which doctorates, habilitations, and professorships were awarded, were used to classify scientists into three academic ranks: the period between achieving doctoral degree and the habilitation degree (if awarded) was regarded as a period of assistant professorship; the period between achieving the habilitation degree and the professorship title (if awarded) was regarded as a period of associate professorship; and the period following the attainment of the professorship title was regarded as a period of full professorship.

We created individual lifetime biographical profiles and individual lifetime publication profiles for every scientist in our sample ($N = 16{,}083$). Biographical profiles included relevant dates in academic careers, and publication profiles included lifetime publication and citation metadata. By combining publication data and biographical data, we were able to allocate every publication to the appropriate stage of the academic career and calculate prestige-normalized individual productivity for any period of time for every scientist in our dataset.

## Methodology

### *Prestige-normalized research productivity*

In our journal prestige-normalized approach (applied for the first time in Kwiek and Roszka 2023), using the Scopus CiteScore percentile ranks, articles published in prestigious journals were given more weight in individual productivity than those published in less prestigious journals within each discipline. Our approach reflects the general idea that articles published in high-impact journals require, on average, more scholarly effort and lead to, on average, greater effects on the scholarly community. Counting all publications evenly tends to disregard vastly different individual scholarly efforts while submitting to a highly stratified system of academic journals (Hammarfelt 2017; Shibayama and Baba 2015; Kwiek 2021a). The journal prestige-normalized approach allows for a fair measurement of scholarly effort in STEMM disciplines. Highly selective top journals are discipline-specific, and journal stratification in science plays a powerful role in academic careers, including academic employment, promotions, and access to competitive research funds, especially in STEMM disciplines (see Supplementary Material).

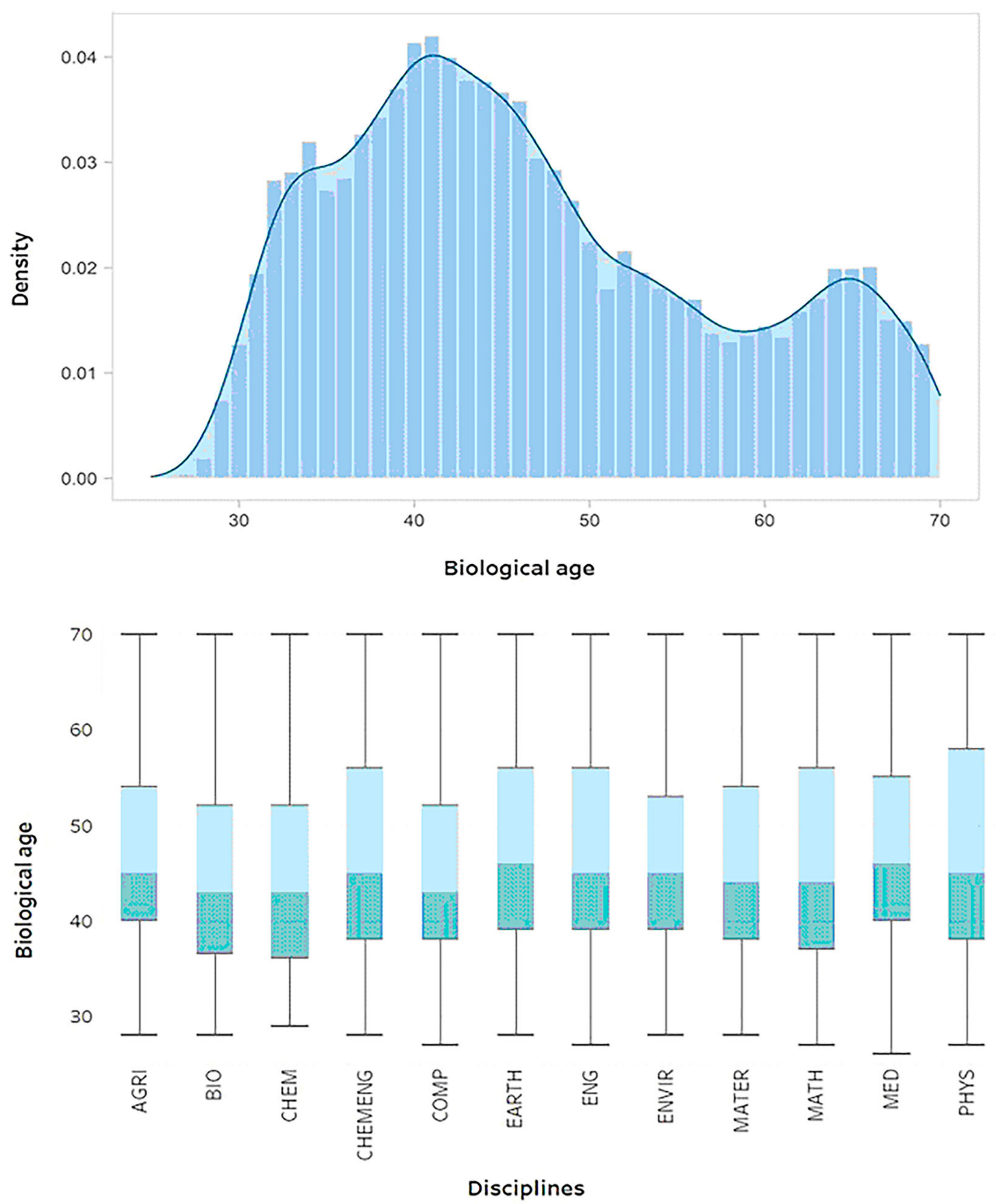

**Figure 1.** Distribution of biological age: kernel density plot (all STEMM academic fields combined) (top panel) and distribution of biological age by discipline (bottom panel, $N = 16{,}083$).

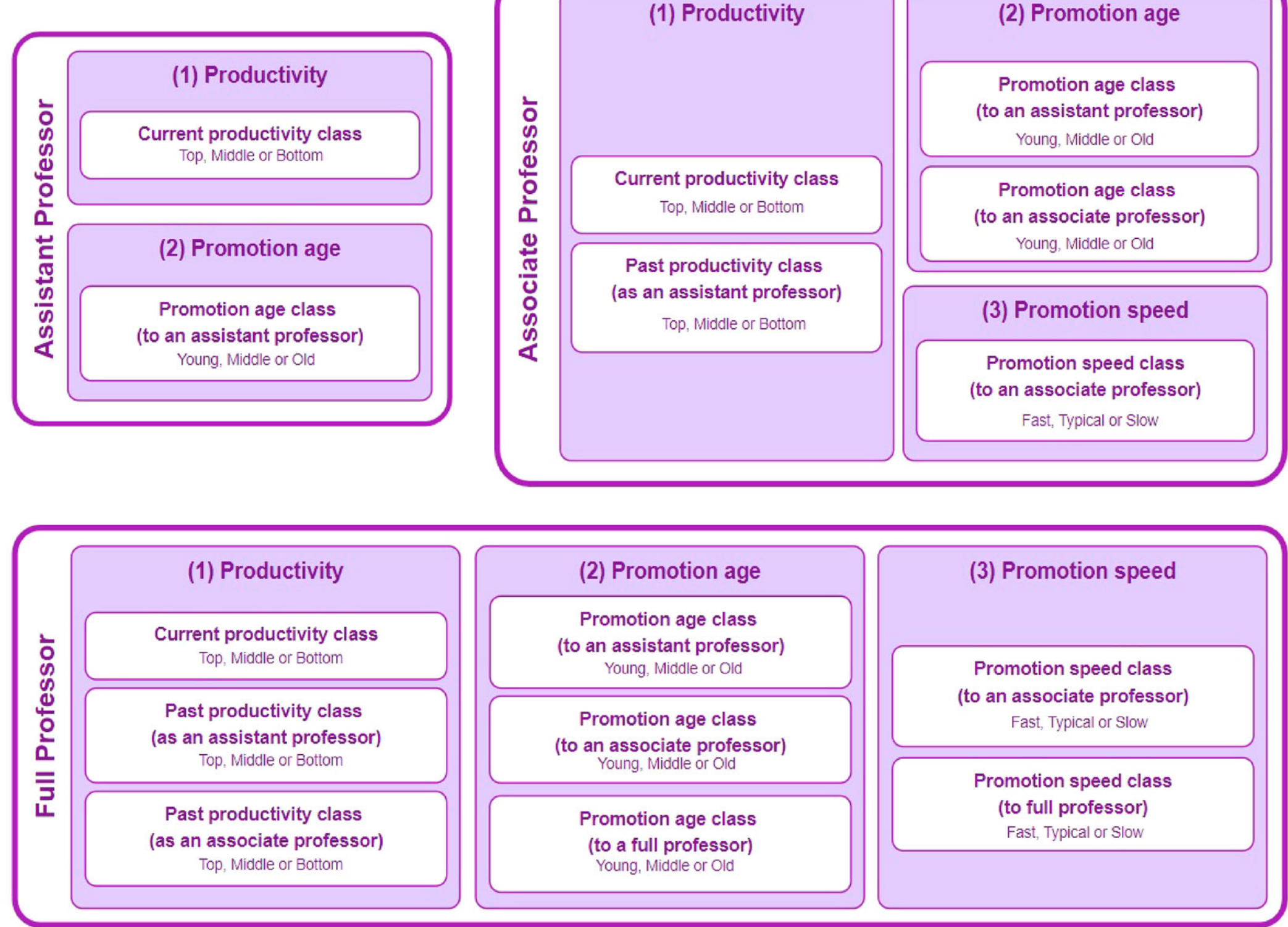

**Figure 2.** Academic career classes. The 20/60/20 classification scheme was used in the analyses. Current (2014–17) and retrospectively constructed productivity (top, middle, and bottom), promotion age (young, middle, and old), and promotion speed (fast, typical, and slow) classes.

### *A classificatory approach to studying academic careers: productivity, promotion age, and promotion speed classes*

Allocating all scientists to three types of academic career classes – productivity, promotion age, and promotion speed – based on the 20/60/20 pattern is key to our research. The higher the rank, the more classes are available (as shown in Figure 2); this is consistent with our retrospective approach, in which every full professor was both associate professor and assistant professor in the past. In the case of full professors, there were three productivity classes (current, past as assistant professor, and past as associate professor), three promotion age classes (age at promotion to an assistant professor, to an associate professor, and to a full professor), and two promotion speed classes (amount of time prior to a promotion to associate professor and prior to full professor).

The productivity classes used were the top 20%, middle 60%, and bottom 20% in a prestige-normalized approach (separately within each of the 12 STEMM disciplines). The promotion age classes of full professors were young, middle, or old associate professors and young, middle, or old full professors. The promotion speed classes of full professors were fast, typical, and slow associate professors, and fast, typical, and slow full professors – that is, the top 20%, middle 60%, and bottom 20%, respectively – in terms of the transition time between subsequent promotions expressed in years. Analogous procedures for constructing current and retrospective academic career classes were applied to associate and assistant professors.

Having identified current professors of different ranks and defined their biographical profiles, we examined their biological ages at previous promotions (promotion age). We also examined the amount of time between promotions (promotion speed). Thus, we examined the current ranks of professors and the distribution of their biological ages at the time of their previous promotions (see Supplementary Material).

## Results

### *Current productivity and past promotion age classes*

In this section, we analyze the current median individual productivity (in the study period of four years, 2014–17) according to the three promotion age classes (young, middle, and old) for three academic ranks. Across all disciplines, the young promotion age class was consistently the most productive, and the old promotion age class was consistently the least productive in terms of median productivity (see the details in Table 2 in Supplementary Material).

Thus, the results showed that the younger the promotion age at all levels, the higher the current productivity. The age classes of past promotions were strongly related to current productivity. The differences were astonishingly and systemically similar across all disciplines. The productivity differential was highest for young promoted and old promoted associate professors and lowest for young promoted and old promoted assistant professors.

For instance, in chemical engineering (CHEMENG), the comparison of the medians showed that young promoted assistant professors had three times higher productivity than old promoted assistant professors (272.8%), young promoted associate professors had five times higher productivity than old promoted associate professors (485.8%), and young promoted full professors had three times higher productivity than old promoted full professors (339.3%, Figure 3).

Although their promotions had occurred in the past, the current productivity of young promoted scientists in terms of promotion age (young promoted class) in all disciplines was clearly higher than the productivity of the other two promotion age classes, especially the old promoted class. These findings were refined using the results of statistical tests, especially pairwise comparisons. Typically, the value of the test statistic in pairwise comparisons was the highest for the old–young promotion class pair. The higher the value of the test statistic, the greater the discrepancy between the distributions. When there was a large discrepancy between distributions, we also found large discrepancies in the characteristics of these distributions (the results of the Kruskal–Wallis test are discussed in Supplementary Material).

### *Current productivity and past promotion speed classes*

In this section, we analyze the current median individual productivity according to the three promotion speed classes (fast, typical, and slow promotion classes) for two academic ranks: associate professor and full professor. Across all disciplines, the results showed that the class that promoted the most quickly in the past (fast) was consistently the most currently productive, and the class that promoted the most slowly in the past (slow) was consistently the least productive in terms of the median (see the details in Table 3 in Supplementary Material).

Thus, the higher the previous promotion speed, the higher the median current productivity. In all disciplines, fast promoted associate professors were, on average, the most productive, and slow promoted associate professors were, on average, the least productive. Similarly, across all disciplines, fast promoted full professors were, on average, the most productive, and slow promoted full professors were, on average, the least productive. The productivity differential was higher for fast and slow promoted associate professors than for fast and slow promoted full professors.

Using the example of physics and astronomy (PHYS), based on the medians, the productivity of fast promoted associate professors was, on average, four times higher (431.4%) than the productivity of slow promoted associate professors, and the productivity of fast promoted full professors was, on average, twice as high (187.7%, Figure 4) as the productivity of slow promoted full professors. The results of the Kruskal–Wallis test are discussed in Supplementary Material.

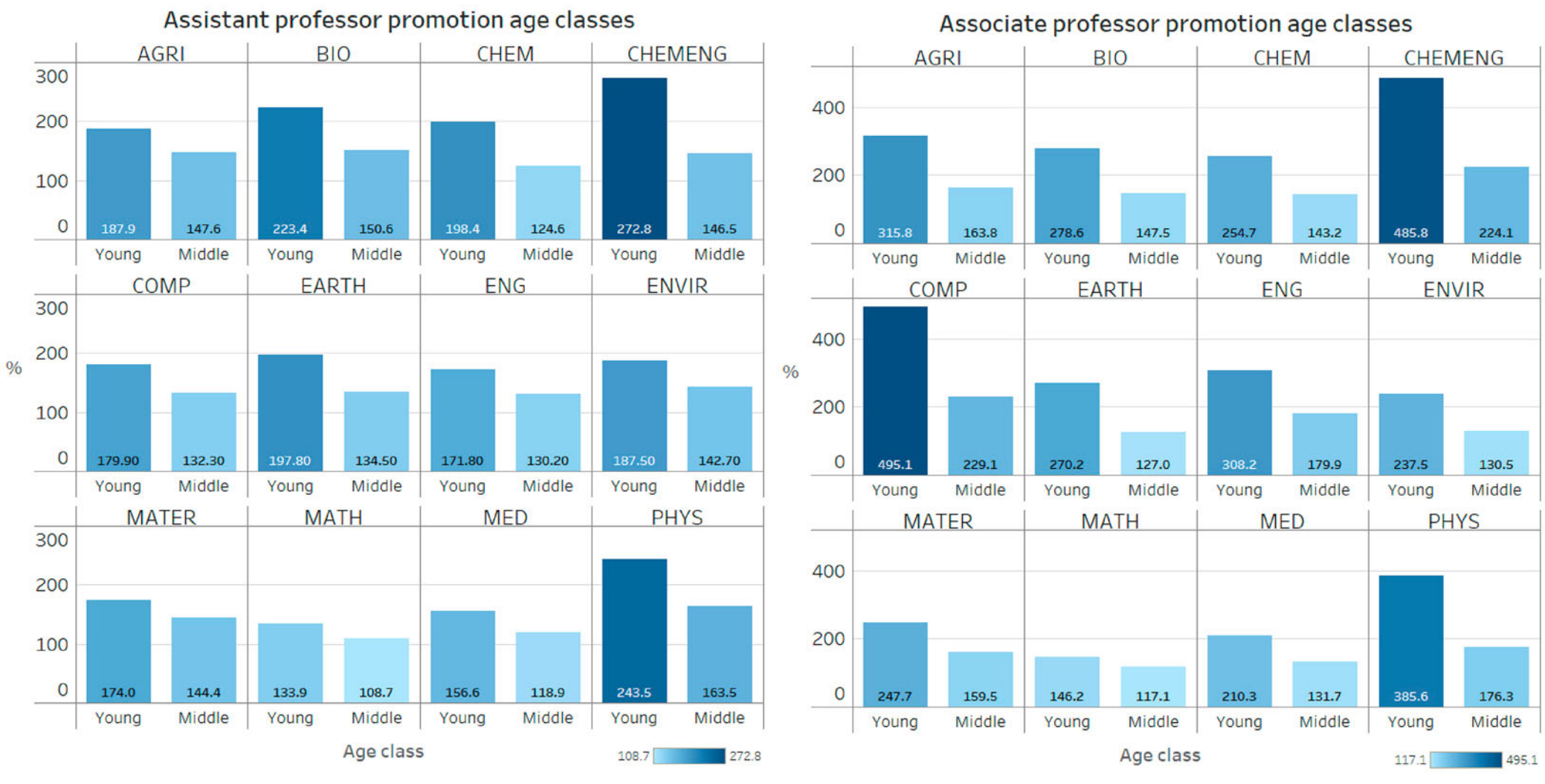

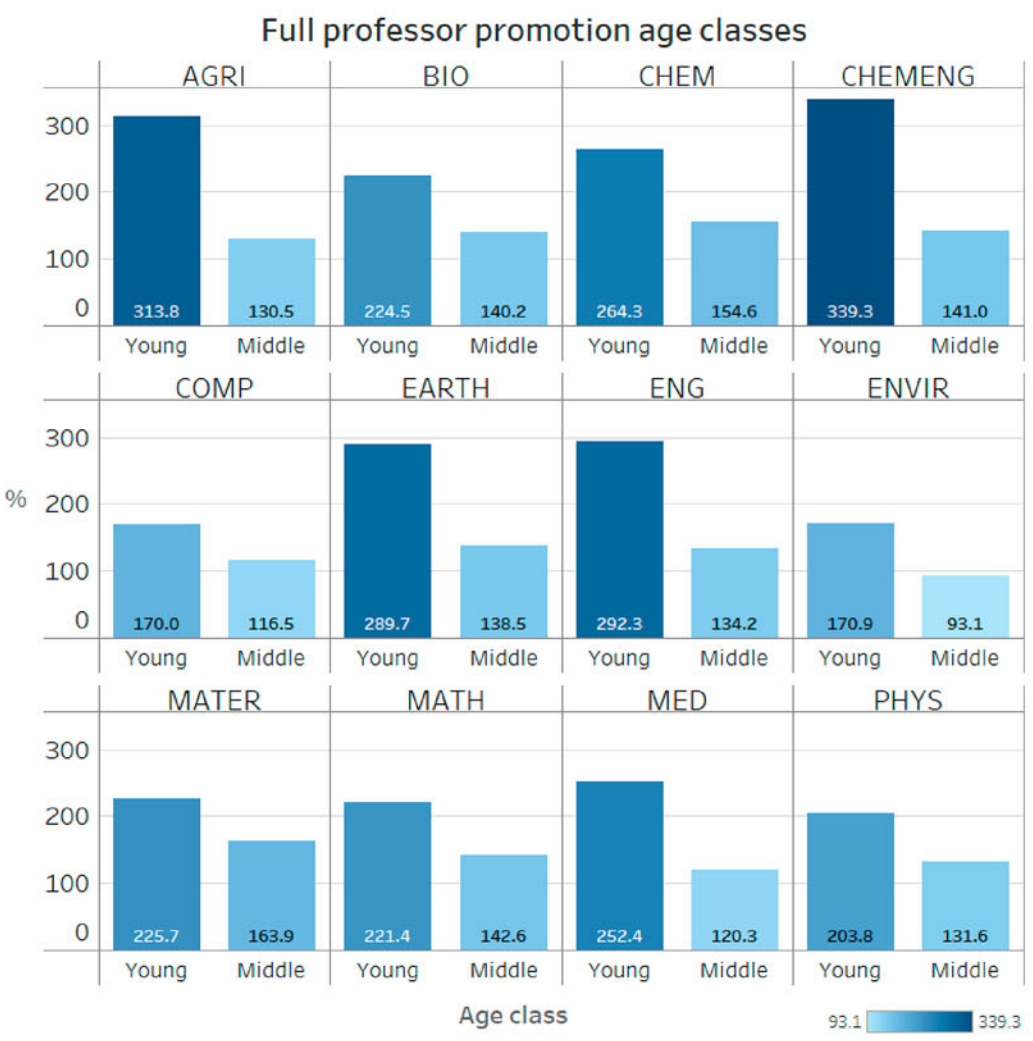

**Figure 3.** Productivity differential for assistant professors (top left panel), associate professors (top right panel) and full professors (bottom panel) between promotion age classes (young, middle, and old promoted) and discipline. Prestige-normalized productivity, full counting, 2014–17. The productivity of the old promotion age class = 100% ($N = 9084$, $N = 4715$, and $N = 2284$, respectively).

## *Logistic regression*

Three logistic regression models were created separately for assistant, associate, and full professors, in which success was defined as membership in the class of the top 20% of the most productive scientists. The predictors increasing the chances of belonging to the current highly productive classes (the top 20%) were sought.

In the models, we used individual-level and organizational-level predictors: gender (binary), biological age and academic age; age of receiving doctorate, habilitation, and full professorship as thresholds for proxies of internationally comparable academic seniority levels ('assistant professorship promotion age', 'associate professorship promotion age', and 'full professorship promotion age'); individual 'average team size (lifetime)' and publication-related individual 'median journal prestige rate (lifetime)'; and 'research intensive institution'. The individual average team size was the median value of the number of collaborators per article in all articles published in a lifetime (see Kwiek 2021b). The individual median journal prestige rate in a scientist's individual publication profile (range: 0–99) was calculated based on all publications in a lifetime. Our variables also included the classes of young promotion assistant, associate, and full professors (promotion age): the classes of fast promotion associate and full professors (promotion time); and the classes of highly productive assistant and associate professors (past productivity). In the models, we included the promotion age factor, the promotion speed factor, and the productivity factor at earlier stages of the academic career, whenever applicable.

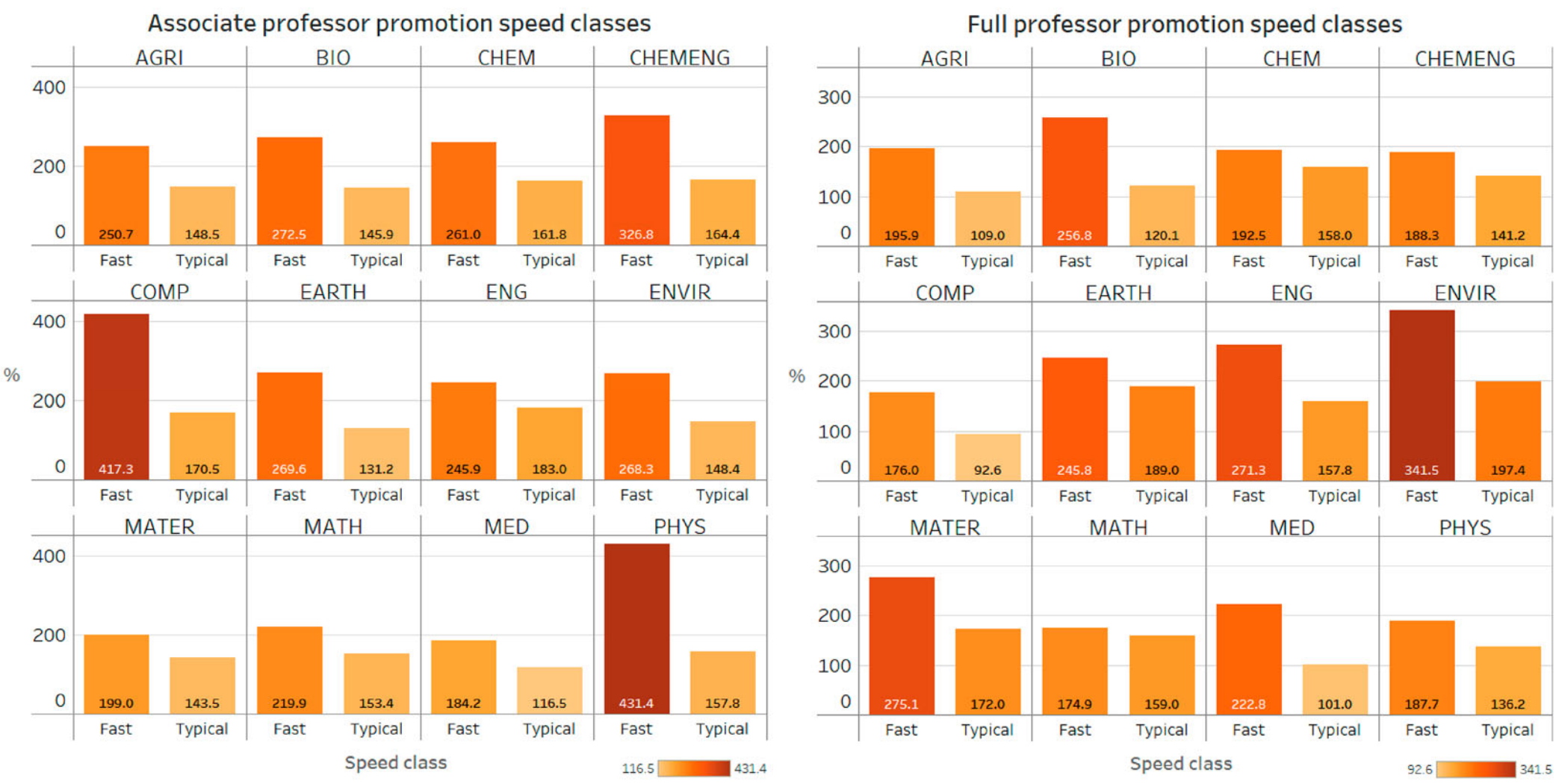

**Figure 4.** Productivity differential for associate professor (left panel) and full professors (right panel) between promotion speed classes (fast, typical, and slow) and discipline. Prestige-normalized productivity, full counting, 2014–17. The productivity of the slow promotion speed class = 100% ($N = 4715$ and $N = 2284$, respectively).

The results of the regression analysis (Table 2) showed that promotion age and promotion speed classes played different roles among the predictors of membership in the current top productivity classes. In the model of full professors (Model 1), the strongest predictor was membership in the class of highly productive associate professors earlier in academic careers, increasing the odds on average by as much as 358%. While gender was statistically insignificant, biological age decreased the odds, and academic age increased the odds (by 4.1% and 2.7%, respectively, with every additional year). In the model of associate professors (Model 2), the results showed a single powerful predictor of membership in the current top productivity class: top productivity in the past as an assistant professor, which increased the odds, on average, by 482%. Six other variables were statistically significant. Membership in the fast associate professor class increased the odds by about one-third (35.6%) on average. In this model, institutional research intensity was significant and decreased the odds by about one-fourth (26.1%; see Supplementary Material for further details).

Overall, the results of the multidimensional analysis of all predictors indicated that the roles of the promotion age and promotion speed classes were not as significant as expected, based on the two-dimensional analysis. The most powerful predictors were membership in high-productivity classes in the past.

**Table 2.** Logistic regression statistics: odds ratio estimates of belonging to the highly productive classes (the top 20%).

| | Model 1: Highly Productive Full Professors $N = 1754$ $R^2 = 0.202$ | | | | Model 2: Highly Productive Associate Professors $N = 4225$ $R^2 = 0.265$ | | | | Model 3: Highly Productive Assistant Professors $N = 9077$ $R^2 = 0.239$ | | | |
|---|---|---|---|---|---|---|---|---|---|---|---|---|
| | | 95% CI for Exp(B) | | | | 95% CI for Exp (B) | | | | 95% CI for Exp (B) | | |
| Model | Exp(B) | LB | UB | Sig. | Exp(B) | LB | UB | Sig. | Exp(B) | LB | UB | Sig. |
| (Intercept) | 0.199 | 0.013 | 2.981 | 0.239 | 0.934 | 0.299 | 2.917 | 0.906 | 1.575 | 0.811 | 3.059 | 0.215 |
| Academic age | 1.027* | 1.003 | 1.051 | 0.028 | 1.051*** | 1.034 | 1.069 | <0.001 | 1.139*** | 1.122 | 1.157 | <0.001 |
| Biological age | 0.959* | 0.928 | 0.991 | 0.010 | 0.932*** | 0.907 | 0.957 | <0.001 | 0.792*** | 0.778 | 0.805 | <0.001 |
| Assistant professorship promotion age | 1.001 | 0.929 | 1.078 | 0.980 | 1.013 | 0.973 | 1.055 | 0.535 | 1.167*** | 1.139 | 1.195 | <0.001 |
| Associate professorship promotion age | 1.096* | 1.019 | 1.178 | 0.026 | 0.996 | 0.964 | 1.030 | 0.810 | – | – | – | – |
| Full professorship promotion age | 0.94 | 0.885 | 0.999 | 0.077 | – | – | – | – | – | – | – | – |
| Male scientists | 1.06 | 0.769 | 1.462 | 0.733 | 1.321** | 1.107 | 1.576 | 0.002 | 1.484*** | 1.323 | 1.664 | <0.001 |
| Young promotion assistant professors class | 0.911 | 0.612 | 1.355 | 0.655 | 1.103 | 0.866 | 1.404 | 0.440 | 1.535*** | 1.334 | 1.765 | <0.001 |
| Young promotion associate professors class | 1.472 | 0.910 | 2.381 | 0.107 | 0.859 | 0.636 | 1.159 | 0.335 | – | – | – | – |
| Young promotion full professors class | 1.399 | 0.868 | 2.257 | 0.158 | – | – | – | – | – | – | – | – |
| Fast promotion associate professors class | 1.276 | 0.838 | 1.943 | 0.247 | 1.356* | 1.029 | 1.787 | 0.033 | – | – | – | – |
| Fast promotion full professors class | 0.913 | 0.621 | 1.341 | 0.660 | - | – | – | – | – | – | – | – |
| Highly productive assistant professors class | 2.831 | 0.489 | 16.376 | 0.408 | 5.824*** | 4.646 | 7.302 | <0.001 | – | – | – | – |
| Highly productive associate professors class | 4.581*** | 2.872 | 7.307 | <0.001 | - | – | – | – | – | – | – | – |
| Average team size (lifetime) | 1.031** | 1.016 | 1.046 | 0.002 | 1.025** | 1.014 | 1.037 | 0.004 | 1.003 | 1.000 | 1.006 | 0.419 |
| Median journal prestige rate (lifetime) | 1.05*** | 1.039 | 1.061 | <0.001 | 1.032*** | 1.028 | 1.036 | <0.001 | 1.028*** | 1.026 | 1.031 | <0.001 |
| Research intensive institution: Rest | 0.916 | 0.691 | 1.216 | 0.538 | 0.739** | 0.619 | 0.883 | 0.001 | 1.015 | 0.901 | 1.144 | 0.805 |

$^*P<0.05$, $^{**}P<0.01$, $^{***}P<0.001$.

## Discussion and conclusions

We constructed individual lifetime biographical profiles and individual lifetime publication profiles for every scientist in our sample of STEMM scientists with doctorates ($N = 16{,}083$). Our research shows the new opportunities provided by merging large-scale national and global datasets to study academic careers by using publication metadata on all Polish articles indexed in Scopus ($N = 935{,}167$).

We used a new methodological approach; instead of traditional productivity, which is based on publication counts (full counting or fractional counting), we used journal prestige-normalized productivity, reflecting differing scholarly efforts and impacts on the academic community. We used a classificatory approach to academic careers, allocating all scientists to different productivity, promotion age, and promotion speed classes based on a 20/60/20 division.

Our research highlighted that rank advancement earlier in academic careers and productivity later in academic careers are strongly linked in ways that have not been discussed in the literature – through the two time-related dimensions of promotion age and promotion speed. Following generally scattered remarks in previous research (e.g. Abramo, D'Angelo, and Murgia 2016; Cole 1979; Cole and Cole 1973; Long, Allison, and McGinnis 1993), our study showed in detail using a large-scale sample of STEMM scientists that assistant, associate, and full professors who were promoted at a young age (the young promoted classes) were, on average, much more productive than assistant, associate, and full professors who were promoted at a later age (the old promoted classes). There have been very rare remarks in previous literature about the links between early promotions (by biological age) and productivity later in academic careers – as opposed to wide previous literature about relationships between age in general and productivity. Studies linking age at subsequent promotions and productivity at later stages of academic careers, and the time between promotions and productivity at later stages, what we develop in this research, do not seem to have been conducted using large datasets in previous research.

The patterns that emerged from our research are surprisingly consistent. First, in all disciplines, scientists in the young promotion age classes (the young promoted, top 20%) were consistently the most productive, and scientists in the old promotion age classes (the old promoted, bottom 20%) were consistently the least productive in all three ranks studied. Thus, current journal prestige-normalized productivity levels (for 2014–17) across all disciplines were strongly related to past promotion age classes. Second, scientists in the fast promotion speed classes (the fast promoted, top 20%) were consistently the most productive, and scientists in the slow promotion speed classes (the slow promoted, bottom 20%) were consistently the least productive. Importantly, this research is not about the young and the old in science, or about age and productivity – but about the young (and fast) promoted vs. the old (and slow) promoted, or about the age at subsequent promotions and productivity.

Thus, the current median productivity was the highest for scientists in both the young (or early) promotion age and fast promotion speed classes across all disciplines. It was the lowest for scientists in the old (or late) promotion age and slow promotion speed classes. However, our results were only partially confirmed by the results of regression analyses in which we examined odds ratio estimates of current membership in top productivity classes. The difference in focus played a role: the focus on median productivity by promotion age classes and by promotion speed classes in a two-dimensional approach and, in contrast, the focus on high research productivity and its predictors in regression analysis. Membership in the promotion age and promotion speed classes emerged as predictors in our regression analysis, but its role was less significant than that of membership in top productivity classes at previous stages of an academic career.

On average, professors in all ranks who were promoted early (at a young age) and fast were substantially more productive than professors in all ranks who were promoted late (at an older age) and slow. However, to speculate about their underlying causes, we need to return to the productivity theories discussed in Theoretical Framework section. Cumulative advantage (Allison and Stewart 1974) and the ‘sacred spark’ (David 1994; Stephan and Levin 1992) theories shed some light on the relationships found in this study. Our findings showed that a small group of highly talented and motivated scientists were consistently highly productive, and they were promoted early and fast, with shorter periods of time between subsequent promotions in a Polish three-rank system. Their career trajectories in each discipline were evident in the individual micro-level data. Thus, for some, the ‘sacred spark’ theory works well and is useful in explaining their productivity success throughout their careers.

For others, cumulative advancement and reinforcement theories were more applicable. According to these theories, we can assume that some scientists who were perceived as more successful by peers were given more financial and reputational resources. They were promoted faster and at a young age. These scientists were successful and recognized by their peers in a system in which promotions were based almost exclusively on publications. Their high productivity was due in part to external stimuli. Promotions to associate professorships at a young age often led to promotions to full professorships at a young age, which was evident in our micro-level data (see coexistence analysis in Supplementary Material).

Finally, the low productivity of scientists in the old promotion age classes and slow promotion speed classes tended to continue throughout their careers. Being older and slower in receiving promotions negatively influences the perceptions of colleagues in research grant panels and peers in their own disciplines. In their case, external awards traditionally accompanying early promotions and fast promotions – peer recognition, access to grants, and higher pay – do not reinforce a sustained focus on research. For scientists in these two comparatively disadvantaged promotion classes (the old promoted and the slow promoted), promotion to associate professorship is achieved. A select few may even be promoted to full professors without sustaining high productivity over time. However, external awards come too late to be effective as external stimuli that promote high productivity.

The mechanism that possibly explains why scientists with young promotions (and fast promotions) are far more productive than those with late promotions (and slow promotions) is similar to the traditional mechanism explaining high and low productivity, except that seeking recognition by publishing, in the Polish case, is replaced by seeking recognition by publishing for successful promotions. The mechanism is consistent with older formulations and therefore our work can also be regarded as a measurement contribution, supporting previous conclusions from the sociology of science while working with much larger observation numbers compared with previous studies.

Our study refers directly to the traditional patterns in the operation of science and the ways in which its operations differentiate careers by replacing the focus on publications with the focus on promotions obtained through publications. The mechanisms discussed for recognition through publications seem to work, in the Polish case, for recognition through promotions. Our research contributes to productivity research in academic profession studies, highlighting the role of path dependence in academic careers. The findings of our study indicate that high productivity is strongly associated with promotions at a young age (and fast rank advancement), and low productivity is associated with promotions at an older age (and slow rank advancement).

What are the implications for university administrators or policymakers? The average distribution of subsequent promotions is reflected in the average distribution of productivity. Some scientists move up the academic ladder fast, and receive their promotions early, and research expectations from them should always be high. Raising the bar for them seems a must. However, universities are also populated by scientists who move up the ladder slow and receive their promotions late or never. Based on our micro-level data, research expectations from the old and slow promotion classes of scientists should be lowered; both before and after promotions, they tend to be low performers. A policy lesson at both institutional and national levels is that, at some point, their energy should be directed more towards teaching or administration as their average chances for high productivity seem marginal. A wider national policy lesson is that the model of strict publication-based requirements for promotions may be inefficient for the system as a whole, although it may still work well in research intensive universities. There are many scientists who will never be productive and therefore less stringent rules, and leaving more power to institutions in rank advancement, could be a viable policy option for the future.

Finally, our study has some limitations. First, its limitation is 'success bias.' Only successful scientists were examined, that is, those who were continuously recorded in the national science system as receiving promotions and publishing research. However, the selection bias in Poland is smaller than in other science systems – as scientists age, both the most productive and the least productive, they tend to stay in academe. Second, data on real scientists in the national registry were combined with data on individual Scopus author IDs collected from Scopus, with a possible error between 'real individuals' and their IDs in a global indexing system (marginal due to the role of Scopus as a provider of data on Polish reforms). Third, major global bibliometric datasets have linguistic, geographic, and disciplinary biases that have been discussed for many years (see e.g. Boekhout, van der Weijden, and Waltman 2021).

## Acknowledgments

We are grateful to Lukasz Szymula from the Poznan Team for his assistance and comments. We gratefully acknowledge the thoughtful comments of the reviewers.

## Disclosure statement

No potential conflict of interest was reported by the author(s).

## Funding

This work was supported by Ministry of Education and Science, Poland [grant number NdS/529032/2021/2021].

## ORCID

*Marek Kwiek* http://orcid.org/0000-0001-7953-1063
*Wojciech Roszka* http://orcid.org/0000-0003-4383-3259